\documentclass[english,aps,manuscript,pra]{revtex4}
\usepackage[T1]{fontenc}
\usepackage[latin9]{inputenc}
\usepackage{xcolor}
\usepackage{amsmath}
\usepackage{amssymb}
\usepackage{esint}
\PassOptionsToPackage{normalem}{ulem}
\usepackage{ulem}

\makeatletter

\providecolor{lyxadded}{rgb}{0,0,1}
\providecolor{lyxdeleted}{rgb}{1,0,0}

\newcommand{\lyxdeleted}[3]{{\color{lyxdeleted}\sout{#3}}}

\@ifundefined{textcolor}{}
{%
 \definecolor{BLACK}{gray}{0}
 \definecolor{WHITE}{gray}{1}
 \definecolor{RED}{rgb}{1,0,0}
 \definecolor{GREEN}{rgb}{0,1,0}
 \definecolor{BLUE}{rgb}{0,0,1}
 \definecolor{CYAN}{cmyk}{1,0,0,0}
 \definecolor{MAGENTA}{cmyk}{0,1,0,0}
 \definecolor{YELLOW}{cmyk}{0,0,1,0}
 }

\usepackage{epic}\usepackage{curves}

\usepackage{babel}

\usepackage{babel}

\makeatother

\usepackage{babel}
\begin{document}

\title{Unambiguous discrimination of extremely similar states by a weak
measurement}

\author{$Chang\, Qiao^{\star}$, $Shengjun\, Wu^{\dagger}$ and $Zeng-Bing\, Chen^{\ddagger}$}

\affiliation{Hefei National Laboratory for Physical Sciences at Microscale and
Department of Modern Physics, University of Science and Technology
of China, Hefei, Anhui 230026, China}

\affiliation{$^{*}qch19@mail.ustc.edu.cn$, $^{\dagger}shengjun@ustc.edu.cn$,
$^{\ddagger}zbchen@ustc.edu.cn$}
\begin{abstract}
In this paper, we propose a method to discriminate two extremely similar
quantum states via a weak measurement. For the two states with equal
prior probabilities, the optimum discrimination probability given
by Ivanovic-Dieks-Peres limit can be achieved by our protocol with
an appropriate choice of the interaction strength. However, compared
with the conventional method for state discrimination, our approach
shows the advantage of error-tolerance by achieving a better ratio
of the success probability to the probability of error. 
\end{abstract}
\maketitle
Since 1980's, the quantum state discrimination problem has been a
subject of active investigation \cite{1,3,4,key-10-1,key-12-1,key-13-1,key-15-1}.
As a given quantum system can be prepared in one of two nonorthogonal
states, from the observed statistical properties we can not determine
the state of the system with certainty. However, we may be able to
use some prior information to determine, at least to some extent,
the state. Optimal measurements have been found to identify the state
with minimal error \cite{29,30,31,32,33,34,37,38,39,40,41}, to discriminate
the states unambiguously \cite{key-53,key-54,key-55,key-56,key-57,key-58,key-59,key-60,key-61,key-62,key-63,key-64},
and to determine the state with the maximum level of confidence \cite{81,83,84,85}. 

For an unambiguous discrimination, any physically realizable measurement
can be described by a probability operator measure, which is known
as a positive operator-valued measure (POVM). Suppose the system is
prepared in one of the two pure states $|\psi_{1}\rangle$ and $|\psi_{2}\rangle$
with prior probabilities $p_{1}$, $p_{2}$. The set of POVM operators
is denoted as $\{\hat{\pi}_{1},\hat{\pi}_{2},\hat{\pi}_{?}\}$ where
$\hat{\pi}_{?}=I-\hat{\pi}_{1}-\hat{\pi}_{2}$. The word {}``unambiguous''
means that $\langle\psi_{1}|\hat{\pi}_{2}|\psi_{1}\rangle=\langle\psi_{2}|\hat{\pi}_{1}|\psi_{2}\rangle=0$.
Thus, when outcome $1$ that is associated with operator $\hat{\pi}_{1}$
is obtained, we can say for sure that the state of the system was
$|\psi_{1}\rangle$, when outcome $2$ that is associated with operator
$\hat{\pi}_{2}$ occurs, we will know that the state was $|\psi_{2}\rangle$
with certainty. For equal prior probabilities, $p_{1}=p_{2}$, Ivanovic\textendash{}Dieks\textendash{}Peres
(IDP) limit gives the maximal success probability for the unambiguous
discrimination 
\begin{equation}
p_{max}=1-|\langle\psi_{1}|\psi_{2}\rangle|.\label{eq:idp}
\end{equation}

In this paper, we use another approach, and perform state discrimination
via a weak measurement \cite{Aharonov1,Aharonov 2,Klaus-1,Wu-1,ah3,Wu-2},
which has been widely used in signal amplification \cite{feedback 1,m weakv,phase l,po weakv,prescion,wu zu,s weakv}.
In a typical weak measurement, the initial state of a system is prepared
in $|\psi_{in}\rangle$, after a weak coupling with an apparatus,
a postselection $|\psi_{f}\rangle$ is performed on the system. Conditional
on the successful postselection, we could observe a large shift of
the pointer (apparatus) position or momentum, which is usually proportional
to the weak value \cite{Joza,key-27,m weakv,po weakv,s weakv,Wu Li,pwc}.
A weak measurement could achieve a signal amplification and a better
signal-to-noise ratio \cite{prescion,wu zu,phase l}. In the following,
we use the advantage of weak measurement in signal amplification and
develop a way of state discrimination with a much more robust error-tolerance.
We shall discuss the unambiguous discrimination of two extremely similar
states via weak measurement.

Assume qubit $A$ is prepared in one of the two pure states: $\rho_{A1}=|\psi_{1}\rangle\langle\psi_{1}|$
and $\rho_{A2}=|\psi_{2}\rangle\langle\psi_{2}|$, with 
\begin{equation}
\begin{aligned}|\psi_{1}\rangle= & \frac{1}{\sqrt{2}}|0\rangle_{x}-\frac{1}{\sqrt{2}}|1\rangle_{x}\\
|\psi_{2}\rangle= & \frac{\eta+\frac{1}{\sqrt{2}}}{\sqrt{1+2|\eta|^{2}}}|0\rangle_{x}+\frac{\eta-\frac{1}{\sqrt{2}}}{\sqrt{1+2|\eta|^{2}}}|1\rangle_{x},
\end{aligned}
\label{eq:f1}
\end{equation}
where $|0\rangle_{x}$ and $|1\rangle_{x}$ are eigenstates of $\hat{\sigma}_{x}$.
It is very hard to discriminate $|\psi_{1}\rangle$ from $|\psi_{2}\rangle$
when $|\langle\psi_{1}|\psi_{2}\rangle|$ is large, namely, when $|\eta|$
is small, as $|\langle\psi_{1}|\psi_{2}\rangle|=\frac{1}{\sqrt{1+2|\eta|^{2}}}$.
From Eq. \eqref{eq:idp}, we know the success probability of discriminating
states $|\psi_{1}\rangle$ and $|\psi_{2}\rangle$ has the following
upper bound 
\begin{equation}
\begin{aligned}p_{max} & =1-\frac{1}{\sqrt{1+2|\eta|^{2}}}\end{aligned}
.\label{eq:op}
\end{equation}

Now, we discuss how to distinguish the two states of qubit $A$ via
a weak measurement. For this purpose, we introduce another qubit system
$B$ as the pointer system (measuring device), and introduce an interaction
Hamiltonian

\begin{equation}
\hat{H}_{int}=g\hbar\hat{\sigma}_{xA}\otimes\hat{\sigma}_{xB}\delta(t-t_{0}),\label{eq:xint-1}
\end{equation}
where $g$ denotes the interaction strength $g\in[0,\pi]$. Suppose
the initial state of qubit $B$ is $|\phi\rangle=|0\rangle_{z}$,
the overall initial state of both qubits is $\left|\psi_{i}\right\rangle \left|0\right\rangle _{z}$
($i=1,2$). After the coupling between qubits $A$ and $B$, the overall
state is given as $|\Psi_{i}\rangle=exp[-ig\hat{\sigma}_{xA}\otimes\hat{\sigma}_{xB}]|\psi_{i}\rangle|\phi\rangle$,
($i=1,2$). Next, we postselect qubit A onto $|\psi_{f}\rangle=\frac{1}{\sqrt{2}}|0\rangle_{x}+\frac{1}{\sqrt{2}}|1\rangle_{x}$.
If the postselection is successful, the final state of qubit B is
$|\phi_{i}^{\prime}\rangle=\langle\psi_{f}|\Psi_{i}\rangle/|\langle\psi_{f}|\Psi_{i}\rangle|$,
where 
\begin{equation}
\begin{aligned}|\phi_{1}^{\prime}\rangle & =|1\rangle_{z}\\
|\phi_{2}^{\prime}\rangle & =\frac{\sqrt{2}\eta^{*}\mathrm{cos}g|0\rangle_{z}-i\mathrm{sin}g|1\rangle_{z}}{\sqrt{2|\eta|^{2}\mathrm{cos^{2}}g+\mathrm{sin^{2}}g}}.
\end{aligned}
\label{eq:fi12}
\end{equation}
We denote $\rho_{B1}=|\phi_{1}^{\prime}\rangle\langle\phi_{1}^{\prime}|$
and $\rho_{B2}=|\phi_{2}^{\prime}\rangle\langle\phi_{2}^{\prime}|$. 

As the discrimination of $|\phi^{\prime}\rangle_{1}$ and $|\phi^{\prime}\rangle_{2}$
would attain the original goal of discriminating $|\psi_{1}\rangle$
and $|\psi_{2}\rangle$, the spirit of our protocol is to replace
the discrimination of states $|\psi_{1}\rangle$ and $|\psi_{2}\rangle$
of qubit $A$ by discriminating the states $|\phi^{\prime}\rangle_{1}$
and $|\phi^{\prime}\rangle_{2}$ of qubit $B$. From Eq. \eqref{eq:fi12},
it is easy to see that the fidelity of $|\phi_{1}^{\prime}\rangle$
and $|\phi_{2}^{\prime}\rangle$ is less than the fidelity of $|\psi_{1}\rangle$
and $|\psi_{2}\rangle$ when $g$ is small. Thus the distinguishability
can be improved in our protocol. 

We define $\rho_{1}^{\prime}$ and $\rho_{2}^{\prime}$ as the density
matrices of qubit $A$ after the coupling, corresponding to the two
initial states $|\psi_{1}\rangle$ and $|\psi_{2}\rangle$. Then the
prior probabilities $\lambda_{1}$ and $\lambda_{2}$ of $|\phi_{1}^{\prime}\rangle$
and $|\phi_{2}^{\prime}\rangle$ are determined by the probabilities
for a successful postselection $|\psi_{f}\rangle$, i.e., 
\begin{equation}
\begin{aligned}\lambda_{1} & =\langle\psi_{f}|\rho_{1}^{\prime}|\psi_{f}\rangle=\frac{2|\eta|^{2}\mathrm{cos^{2}}g+\mathrm{sin^{2}}g}{1+2|\eta|^{2}}\\
\lambda_{2} & =\langle\psi_{f}|\rho_{2}^{\prime}|\psi_{f}\rangle=\mathrm{sin^{2}}g.
\end{aligned}
\label{eq:12}
\end{equation}
We are going to consider the case with equal prior probabilities $\lambda_{1}\approx\lambda_{2}$,
or $\lambda_{i}\gg|\lambda_{1}-\lambda_{2}|$ $(i=1,2)$, which is
equivalent to the condition 
\begin{equation}
g\approx\frac{\pi}{4}\mathrm{\: or}\:|\eta|\ll g.\label{eq:condition}
\end{equation}
The average success probability for postselection is $\frac{1}{2}(\lambda_{1}+\lambda_{2})$,
therefore the maximal achievable probability of discrimination via
our strategy is 
\begin{equation}
\begin{aligned}p & =\frac{1}{2}(\lambda_{1}+\lambda_{2})(1-|\langle\phi_{1}^{\prime}|\phi_{2}^{\prime}\rangle|)\\
 & =\frac{|\eta|^{2}+\mathrm{sin^{2}}g}{1+2|\eta|^{2}}(1-\frac{|\mathrm{sin}\, g|}{\sqrt{2|\eta|^{2}\mathrm{cos^{2}}g+\mathrm{sin^{2}}g}}).
\end{aligned}
\label{eq:pus}
\end{equation}
In the following, we shall show that the upper bound $p_{max}$ in
\eqref{eq:idp} can be reached approximately via our strategy with
weak measurement when $g$ and $|\eta|$ satisfy a certain condition.

To ensure $\lambda_{1}\approx\lambda_{2}$, either $g\approx\frac{\pi}{4}$
or $|\eta|\ll g$ must be satisfied, both cases are discussed separately
as follows. When $g\approx\frac{\pi}{4}$, we have $p=\frac{1}{2}(1-\frac{1}{\sqrt{2|\eta|^{2}+1}})=\frac{1}{2}p_{max}$,
which means the maximum cannot be reached, so this case will not be
considered any more. When $|\eta|\ll g$, then we obtain $\lambda_{1}\approx\lambda_{2}=sin^{2}g.$
Then from Eq.\eqref{eq:pus}, the overall probability of our protocol
is given as 
\begin{equation}
\begin{aligned}p & =|\eta|^{2}\mathrm{cos^{2}}g\end{aligned}
.\label{eq:oup}
\end{equation}
The upper bound for the discrimination probability is $p_{max}=|\eta|^{2}$
when $|\psi_{1}\rangle$ and $|\psi_{2}\rangle$ are extremely similar
($|\eta|\ll g$), when the interaction strength $g$ is small, but
$g\gg|\eta|$, the success probability of discrimination via our protocol
reaches the optimal probability $p\rightarrow p_{max}$. Generally
speaking, for the discrimination of two extremely similar states,
with an appropriate choice of the interaction strength, i.e, $|\eta|\ll g\ll1$,
our protocol achieves the optimal discrimination probability. In the
following, we shall show the advantage of our protocol in presence
of imperfections for operations.

Let us rephrase the derivation above with density matrix. The two
possible initial states of qubits $A$ are $\rho_{A1}=\frac{1}{2}(I+\overrightarrow{k_{A1}}\cdot\hat{\sigma})$
and $\rho_{A2}=\frac{1}{2}(I+\overrightarrow{k_{A2}}\cdot\hat{\sigma})$,
the state of qubit $B$ is $\rho_{B}=\frac{1}{2}(I+\overrightarrow{k_{B}}\cdot\hat{\sigma})$,
the interaction evolution is $U=exp[-ig(\overrightarrow{n}\cdot\hat{\sigma}_{A})\otimes(\overrightarrow{n}\cdot\hat{\sigma}_{B})]$,
and the postselection for qubit $A$ is $\Pi_{f_{A}}=\frac{1}{2}(I+\overrightarrow{f_{A}}\cdot\hat{\sigma})$,
where $\overrightarrow{k_{A1}}$ , $\overrightarrow{k_{A2}}$, $\overrightarrow{k_{B}}$,
$\overrightarrow{f_{A}}$ and $\overrightarrow{n}$ are all unit vectors.
After the postselection, two possible final states of qubit $B$ are
given as 
\begin{equation}
\begin{aligned}\rho_{B1}=\frac{1}{2}(I+\overrightarrow{k_{B1}}\cdot\hat{\sigma})= & \frac{Tr_{A}(\Pi_{f_{A}}U\rho_{A1}\otimes\rho_{B}U^{\dagger})}{Tr(\Pi_{f_{A}}\rho_{A})}\\
\rho_{B2}=\frac{1}{2}(I+\overrightarrow{k_{B2}}\cdot\hat{\sigma})= & \frac{Tr_{A}(\Pi_{f_{A}}U\rho_{A2}\otimes\rho_{B}U^{\dagger})}{Tr(\Pi_{f_{A}}\rho_{A})}.
\end{aligned}
\label{eq:bbb}
\end{equation}
The vectors $\overrightarrow{k_{B1}}$ and $\overrightarrow{k_{B2}}$
can be expressed in terms of $\overrightarrow{k_{B}}$ , $\overrightarrow{n}$
and $\overrightarrow{n}\times\overrightarrow{k_{B}}$: 
\begin{equation}
\begin{aligned}\overrightarrow{k_{Bi}}= & c_{1}\overrightarrow{k_{B}}+c_{2}\overrightarrow{n}+c_{3}(\overrightarrow{n}\times\overrightarrow{k_{B}}),\:(i=1,2)\, with\\
c_{1}= & \frac{\alpha_{1}-\alpha_{3}}{\alpha_{1}+\alpha_{2}(\overrightarrow{n}\cdot\overrightarrow{k_{B}})+\alpha_{3}}\\
c_{2}= & \frac{\alpha_{2}+\alpha_{3}(\overrightarrow{n}\cdot\overrightarrow{k_{B}})}{\alpha_{1}+\alpha_{2}(\overrightarrow{n}\cdot\overrightarrow{k_{B}})+\alpha_{3}}\\
c_{3}= & \frac{\alpha_{4}}{\alpha_{1}+\alpha_{2}(\overrightarrow{n}\cdot\overrightarrow{k_{B}})+\alpha_{3}},
\end{aligned}
\label{eq:xis}
\end{equation}
where the four coefficients $\alpha_{1}$, $\alpha_{2}$, $\alpha_{3}$
and $\alpha_{4}$ determined by the operations on system $A$ are
given as $\alpha_{1}=\mathrm{cos^{2}}g(1+\overrightarrow{f_{A}}\cdot\overrightarrow{k_{Ai}})$,
$\alpha_{2}=2\mathrm{sin}g\mathrm{cos}g\overrightarrow{f_{A}}\cdot(\overrightarrow{n}\times\overrightarrow{k_{Ai}})$,
$\alpha_{3}=\mathrm{sin^{2}}g[1+(\overrightarrow{n}\cdot\overrightarrow{k_{Ai}})(\overrightarrow{f_{A}}\cdot\overrightarrow{n})-\overrightarrow{f_{A}}\cdot(\overrightarrow{n}\times\overrightarrow{k_{Ai}}\times\overrightarrow{n})]$
and $\alpha_{4}=2(\overrightarrow{n}\cdot\overrightarrow{k_{Ai}}+\overrightarrow{f_{A}}\cdot\overrightarrow{n})\mathrm{sin}g\mathrm{cos}g$.
Eq.\eqref{eq:xis} expresses the same relations as Eq.\eqref{eq:fi12}
in terms of vectors on the Bloch sphere. As our calculation becomes
vector synthesis in the three-dimensional space, naturally, the imperfections
for the operations can be considered as some deviations for these
unit vectors $\overrightarrow{k_{B}}$, $\overrightarrow{f_{A}}$
and $\overrightarrow{n}$. 

However, we shall have one direction which is well-defined and concern
with relative deviations for the vectors along the other directions,
namely, we will not consider a globle deviation for $\overrightarrow{k_{B}}$,
$\overrightarrow{f_{A}}$ and $\overrightarrow{n}$. Thus, coordinate
$\overrightarrow{Z}$ are considered well-defined, the initial apparatus
states of qubit $B$ is described as $\overrightarrow{k_{B}}=\overrightarrow{Z}$,
the Bloch vector of the postselection on qubit $A$ is $\overrightarrow{f_{A}}=\overrightarrow{Z}$.
Compared with the one-party operation, a two-party operation could
be more imperfect. So the error for the final states of qubit $B$
is originated from the imperfect interaction evolution, which can
be described as a deviation $\overrightarrow{\delta_{n}}$ for this
two-party operation vector $\overrightarrow{n}$ i.e., $\overrightarrow{n}=\sqrt{1-|\overrightarrow{\delta_{n}}|^{2}}\overrightarrow{X}+\overrightarrow{\delta_{n}}\, with\, X\cdot\overrightarrow{\delta_{n}}=0$.
Without loss of generality, we define coordinate$\overrightarrow{Y}$
as the direction of the difference between the two vectors $\overrightarrow{k_{A1}}$
and $\overrightarrow{k_{A2}}$. Thus we can write $\overrightarrow{k_{A1}}=-\overrightarrow{Z}$
and $\overrightarrow{k_{A2}}=-(\sqrt{1-\epsilon^{2}})\overrightarrow{Z}+\epsilon\overrightarrow{Y},\, with\,1\gg\epsilon\geq0$.
Assume$|\overrightarrow{\delta_{n}}|$ is of the same order of the
magnitude of $\epsilon$. With the appropriate choice of the interaction
strength $1\gg g\gg\epsilon$, we simplify Eq.\eqref{eq:xis} by keeping
only the terms to the first order of $\epsilon$, $|\overrightarrow{\delta_{n}|}$
and the second order of $g$: 
\begin{equation}
\begin{aligned}\overrightarrow{k_{B1}} & =-\overrightarrow{Z}+\mathcal{O}(|\overrightarrow{\delta_{n}}|^{2})\\
\overrightarrow{k_{B2}} & =-\sqrt{1-(\frac{2\epsilon}{g})^{2}}\overrightarrow{Z}+\frac{2\epsilon}{g}\overrightarrow{X}+\mathcal{O}(|\overrightarrow{\delta_{n}}|^{2}).
\end{aligned}
\label{eq:b12}
\end{equation}
Due to the weak interaction strength ($g\ll1$), the fidelity of the
two states $\rho_{B1}=\frac{1}{2}(I+\overrightarrow{k_{B1}}\cdot\hat{\sigma})$
and $\rho_{B2}=\frac{1}{2}(I+\overrightarrow{k_{B2}}\cdot\hat{\sigma})$
is much smaller than that of the original pair $\rho_{A1}=\frac{1}{2}(I+\overrightarrow{k_{A1}}\cdot\hat{\sigma})$
and $\rho_{A2}=\frac{1}{2}(I+\overrightarrow{k_{A2}}\cdot\hat{\sigma})$,
namely, $|\overrightarrow{k_{B1}}-\overrightarrow{k_{B2}}|\gg|\overrightarrow{k_{A1}}-\overrightarrow{k_{A2}}|$.
Thus, the difference between the two original states is amplified.
From Eq.\eqref{eq:b12}, deviation $\overrightarrow{\delta_{n}}$
for the two-party operation vector is not amplified up to the second
order $\mathcal{O}(|\overrightarrow{\delta_{n}}|^{2})$. Our tolerance
of the error $\overrightarrow{\delta_{n}}$ is robust and we will
ignore the imperfection $\overrightarrow{\delta_{n}}$ of the two-party
operation in the following. After the weak measurement, the two states
$\rho_{B1}=\frac{1}{2}(I+\overrightarrow{k_{B1}}\cdot\hat{\sigma})$
and $\rho_{B2}=\frac{1}{2}(I+\overrightarrow{k_{B2}}\cdot\hat{\sigma})$
are to be discriminated with a set of POVM operators, we will compare
our protocol with the conventional method for state discrimination.

For an unambiguous discrimination of two pure states, the set of POVM
operators is denoted as $\{\hat{\pi}_{1},\hat{\pi}_{2},\hat{\pi}_{?}\}$
where $\hat{\pi}_{?}=I-\hat{\pi}_{1}-\hat{\pi}_{2}$. We describe
the imperfections of one-party operations $\{\hat{\pi}_{1},\hat{\pi}_{2},\hat{\pi}_{?}\}$
as the deviations of their Bloch vectors. Following the spirit of
the unambiguous discrimination, the Bloch vectors of $\hat{\pi}_{1}$
and $\hat{\pi}_{2}$ are opposite to the Bloch vectors of the two
states respectively. For a conventional method, the two states which
shall be discriminated with a set of POVM operators $\{\hat{\pi}_{A1},\hat{\pi}_{A2},\hat{\pi}_{A?}\}$
are $\rho_{A1}$ and $\rho_{A2}$, while in our protocol with a weak
measurement, the two states which shall be discriminated with another
set of POVM operators $\{\hat{\pi}_{B1},\hat{\pi}_{B2},\hat{\pi}_{B?}\}$
are $\rho_{B1}$ and $\rho_{B2}$, and the four Bloch sphere vectors
of $\rho_{A1}$, $\rho_{A2}$ and $\rho_{B1}$, $\rho_{B2}$ are $\overrightarrow{k_{A1}}=-\overrightarrow{Z}$,
$\overrightarrow{k_{A2}}=-(\sqrt{1-\epsilon^{2}})\overrightarrow{Z}+\epsilon\overrightarrow{Y}$
and $\overrightarrow{k_{B1}}=-\overrightarrow{Z}$, $\overrightarrow{k_{B2}}=-\sqrt{1-(\frac{2\epsilon}{g})^{2}}\overrightarrow{Z}+\frac{2\epsilon}{g}\overrightarrow{X}$.
Because the Bloch vectors along $\overrightarrow{Z}$ are considered
well-defined, we ignore the deviations for the Bloch sphere vectors
of $\hat{\pi}_{A2}$ and $\hat{\pi}_{B2}$ which are supposed to be
along $\overrightarrow{Z}$, and only consider the imperfections of
$\hat{\pi}_{A1}$ and $\hat{\pi}_{B1}$. Due to $\frac{2\epsilon}{g}\ll1$,
the Bloch vectors of these two POVM elements $\hat{\pi}_{A1}$ and
$\hat{\pi}_{B1}$ are also nearly along $\overrightarrow{Z}$, hence,
we describe the deviation for each of the two Bloch vectors as $\overrightarrow{\delta_{f}}$,
a tiny deviation in the plane of $X-Y$ ($\overrightarrow{\delta_{f}}\cdot\overrightarrow{Z}=0$).
Assume $|\overrightarrow{\delta_{f}}|$ are also of\lyxdeleted{Administrator}{Sun Feb 24 23:06:49 2013}{
} the same order of the magnitude of $\epsilon$. Then the set of
POVM elements for the conventional method is 
\begin{equation}
\begin{aligned}\hat{\pi}_{A1}= & \frac{1}{4-\frac{\epsilon^{2}}{4}}[I+(\sqrt{1-(\epsilon+\overrightarrow{\delta_{f}}\cdot\overrightarrow{Y})^{2}-(\overrightarrow{\delta_{f}}\cdot\overrightarrow{X})^{2}}\overrightarrow{Z}-\epsilon\overrightarrow{Y}+\overrightarrow{\delta_{f}})\cdot\hat{\sigma}]\\
\hat{\pi}_{A2}= & \frac{1}{4-\frac{\epsilon^{2}}{4}}(I+\overrightarrow{Z}\cdot\hat{\sigma})\\
\hat{\pi}_{A?}= & I-\hat{\pi}_{1}-\hat{\pi}_{2}.
\end{aligned}
\label{eq:pome}
\end{equation}
Similarly, the set of POVM elements for our protocol is 
\begin{equation}
\begin{aligned}\hat{\pi}_{B1}= & \frac{1}{4-\frac{\epsilon^{2}}{g^{2}}}[I+(\sqrt{1-[\frac{2\epsilon}{g}+(\overrightarrow{\delta_{f}}\cdot\overrightarrow{X})]^{2}-(\overrightarrow{\delta_{f}}\cdot\overrightarrow{Y})^{2}}\overrightarrow{Z}-\frac{2\epsilon}{g}\overrightarrow{X}+\overrightarrow{\delta_{f}})\cdot\hat{\sigma}]\\
\hat{\pi}_{B2}= & \frac{1}{4-\frac{\epsilon^{2}}{g^{2}}}(I+\overrightarrow{Z}\cdot\hat{\sigma})\\
\hat{\pi}_{B?}= & I-\hat{\pi}_{1}-\hat{\pi}_{2}.
\end{aligned}
\label{eq:pome-1}
\end{equation}
Due to these imperfections of the operations, the discrimination can
not be unambiguous any more, but probabilistic. For both of the protocols
above, we define a ratio of the success probability to the probability
of error as $\beta$. For the conventional method, 
\begin{equation}
\begin{aligned}\beta_{A} & =\frac{Tr(\rho_{A1}\hat{\pi}_{A1})+Tr(\rho_{A2}\hat{\pi}_{A2})}{Tr(\rho_{A1}\hat{\pi}_{A2})+Tr(\rho_{A2}\hat{\pi}_{A1})}\\
 & =1+\frac{\epsilon^{2}-\epsilon(\overrightarrow{\delta_{f}}\cdot\overrightarrow{Y})}{\frac{1}{2}|\overrightarrow{\delta_{f}}|^{2}}.
\end{aligned}
\label{eq:ar}
\end{equation}
For our protocol, the ratio is 
\begin{equation}
\begin{aligned}\beta_{B} & =\frac{Tr(\rho_{B1}\hat{\pi}_{B1})+Tr(\rho_{B2}\hat{\pi}_{B2})}{Tr(\rho_{B1}\hat{\pi}_{B2})+Tr(\rho_{B2}\hat{\pi}_{B1})}\\
 & =1+\frac{(\frac{2}{g}\epsilon)^{2}-(\frac{2}{g}\epsilon)(\overrightarrow{\delta_{f}}\cdot\overrightarrow{X})}{\frac{1}{2}|\overrightarrow{\delta_{f}}|^{2}}.
\end{aligned}
\label{eq:br}
\end{equation}
Now we compare $\beta_{A}$ with $\beta_{B}$. Obviously, from Eq.\eqref{eq:ar}
and Eq.\eqref{eq:br}, we have $\beta_{A},\beta_{B}\geq\frac{1}{2}$.
Since $\overrightarrow{\delta_{f}}$ is a random vector in the plane
of $X-Y$, we can gain average values of the ratio $\beta_{A}$ and
the ratio $\beta_{B}$ 
\begin{equation}
\begin{aligned}\text{\ensuremath{\langle}}\beta_{A}\rangle & =1+\frac{2\epsilon^{2}}{|\overrightarrow{\delta_{f}}|^{2}}\\
\text{\ensuremath{\langle}}\beta_{B}\rangle & =1+(\frac{2}{g})^{2}\cdot\frac{2\epsilon^{2}}{|\overrightarrow{\delta_{f}}|^{2}}.
\end{aligned}
\label{eq:av}
\end{equation}
From Eq.\eqref{eq:av}, we easily see that our protocol with a weak
measurement can achieve a much better ratio of the success probability
to the probability of error $\text{\ensuremath{\langle}}\beta_{B}\rangle\gg\text{\ensuremath{\langle}}\beta_{A}\rangle$.
Due to the weak interaction strength ($g\ll1$), we can have $\text{\ensuremath{\langle}}\beta_{B}\rangle\gg1$,
namely, our success probability is much more greater than our probability
of error. Therefore, our discrimination can be still considered unambiguous. 

Conclusion: We propose a protocol via a weak measurement to discriminate
two extremely similar states. Our protocol amplifies the difference
of the original two states, and achieves a much better error-tolerance
(i.e., the ratio of the success probability to the probability of
error), compared with the conventional method. 

Although our protocol relies on a successful postselection which may
constrain the overall successful probability of discrimination, we
prove that, based on an appropriate choice of the interaction strength
$g$ (i.e., $1\gg g\gg|\eta|$), the optimal discrimination probability
predicted by Ivanovic\textendash{}Dieks\textendash{}Peres (IDP) limit
can be reached. In the presence of imperfections in the interaction
and POVM, we find that our ratio of the success probability to the
probability of error is much larger than the ratio of a conventional
method.
\begin{acknowledgments}
This work is supported by the NSFC (Grants No. 11075148 and No.61125502),
CAS, and the National Fundamental Research Program (Grant No. 2011CB921300).\end{acknowledgments}

\end{document}